# Plagiarism Detection on Electronic Text based Assignments using Vector Space Model

MAC Jiffriya  MAC Akmal Jahan
Post Graduate Institute of Science
University of Peradeniya
Peradeniya, Sri Lanka

Roshan G. Ragel
Department of Computer Engineering
University of Peradeniya
Peradeniya, Sri Lanka

*Abstract*— Plagiarism is known as illegal use of others' part of work or whole work as one's own in any field such as art, poetry, literature, cinema, research and other creative forms of study. Plagiarism is one of the important issues in academic and research fields and giving more concern in academic systems. The situation is even worse with the availability of ample resources on the web. This paper focuses on an effective plagiarism detection tool on identifying suitable intra-corpal plagiarism detection for text based assignments by comparing unigram, bigram, trigram of vector space model with cosine similarity measure. Manually evaluated, labelled dataset was tested using unigram, bigram and trigram vector. Even though trigram vector consumes comparatively more time, it shows better results with the labelled data. In addition, the selected trigram vector space model with cosine similarity measure is compared with tri-gram sequence matching technique with Jaccard measure. In the results, cosine similarity score shows slightly higher values than the other. Because, it focuses on giving more weight for terms that do not frequently exist in the dataset and cosine similarity measure using trigram technique is more preferable than the other. Therefore, we present our new tool and it could be used as an effective tool to evaluate text based electronic assignments and minimize the plagiarism among students.

## I. INTRODUCTION

Plagiarism is known as illegal use of others' part of work or whole work as one's own in any field such as art, poetry, literature, cinema, research and other creative things. Plagiarism is one of the growing issues in academic and research field and giving more concern in University System. Plagiarism diminishes one's innovative thinking, creativeness, imagination and improvement of knowledge. And also it is considered as unethical behaviour in a moral society.

In a survey that was conducted on plagiarism in the academic field at the University of California in Berkley, it was shown that the percentage of plagiarism has increased by 74.4 % within four years period (1993 – 1997) [1] and more than 90% of high school students involve in plagiarism [2]. Plagiarism can take place in many circumstances in the field of academics: 1) during programming 2) assignment submission, 3) developing web content 4) publishing news, articles and research work. 70% of students involve in at least little plagiarism and 40% of students committed plagiarism during the assignment submission using "copy paste" method [13]. Nowadays, plagiarism on assignment is seen abundant among university students. Therefore, it is a potential issue during the submission of assignments in a university due to the abundance of electronic resources on the internet. Some students prepare assignments by copying or paraphrasing colleague's assignment and submitting them without any effort. It has become a very common issue among undergraduate students and academics who face difficulties when evaluating the students own work and their creativeness.

There are several different approaches to detect plagiarism on text based documents. Our research focuses on identifying a suitable intra-corpal plagiarism detection approach for text based assignments by comparing unigram, bigram, and trigram of vector space models with cosine measure and tri-gram sequence matching technique with Jaccard measure. It helps us to evaluate assignments and minimize the plagiarism among students when evaluating them.

The rest of the paper is organized as follows. In Section II, we present a literature survey on plagiarism detection and in Section III, we present our methodology used. In Section IV, we present our experimental results and a discussion on the results in Section V and we conclude this paper in Section VI.

## II. LITERATURE SURVEY

In academic field, plagiarism is classified as source code plagiarism and free text plagiarism. In early days, most researchers focused on source code plagiarism and several tools were developed for source code plagiarism such as Plagio Guard [3, 4], JPlag [5], Moss [6], Saxon, Detecta Copius [7], Sherlock [8], Copy/Paste Detector (CPD) and Big Brother [9]. Similarly there are several text based plagiarism detection tools which support to detect only extra-corpal such as Dupli Checker and Article Checker [10] or both such as Plagiarism Checker X, Turnitin and Ferret. In extra-corpal, materials are compared with outside materials such as the internet resources, whereas intra-corpal plagiarism detection is comparing materials within the learning community. Thus original and plagiarized materials are found in same place [11].

Content based approach and stylometry-based approach are the mostly used approaches for analysing of plagiarism on text based documents [22]. Content based approach focuses on semantic features of document while stylometry-based approach focuses on grammatical style of document [21]. Content based approach involves exact matching of string or set of characters [12]. It is useful to identify plagiarism as students "copy & paste" from others documents. It shows false positive plagiarism during the paraphrasing and verbatim copying of document. Stylometry-based approach is used based on the assumption of every author applies unique grammar style. Both content based approach and stylometry based approach for plagiarism detection consist of two main tasks: 1) document representation; and 2) computation of similarity scores. Content based approach is widely used to detect duplicate documents [18] [19]. The content of a document is represented in several methods such as sequence of bag of words, vector space model, tree and graph [20] [22] [23] [24].

The first task of the plagiarism detection process is document representation. Vector space model is a way of representing text based documents generated by product of term to do mathematical computation in the field of information retrieval by applying queries and plagiarism detection. The representation of document is based on assumption of independence relationship among terms and is heavily useful to search nearly duplicated documents [23]. Tree or graph structures also can be used to represent a text document based on the relationship of terms or part of speech. Graph representation helps to maintain inherent structural information of documents and also help to quickly search related documents [20].

The second task in plagiarism detection is the computation of similarity scores between documents. Popular similarity measurements are Jaccard similarity, cosine similarity, overlap similarity, dice similarity, Humming distance and Euclidean distance [19] [16]. In recent work on [17], assignments were clustered by using Euclidean distance and plagiarism within clustered assignments is detected using trigram sequence matching technique with Jaccard similarity to minimize the processing time. Vector based representation of document deploy cosine similarity [23] [15]. They used unigram vector space model to detect nearest neighbour documents with cosine similarity [23]. Brooke and Hirst applied vector space model for intrinsic plagiarism detection [14]. In a recent work [15], vector space model is used to compute similarity between source document and suspicious document using cosine similarity.

In our work, we are going to compare Vector space model using unigram, bigram and trigram techniques with cosine similarity and comparing the best approach with trigram of Jaccard similarity technique.

III. METHODOLOGY

The methodology presented in Fig. 1. follows a number of stages and they are: 1) Data collection and pre-processing of the assignments; 2) Constructing vectors using unigram, bigram and trigram methods; 3) Applying cosine similarity measure for each method; 4) Constructing list of sequence of trigram for each assignment; and 5) Comparing each pair of assignment using Jaccard similarity measure. Let us look at details of each stage in the rest of this section.

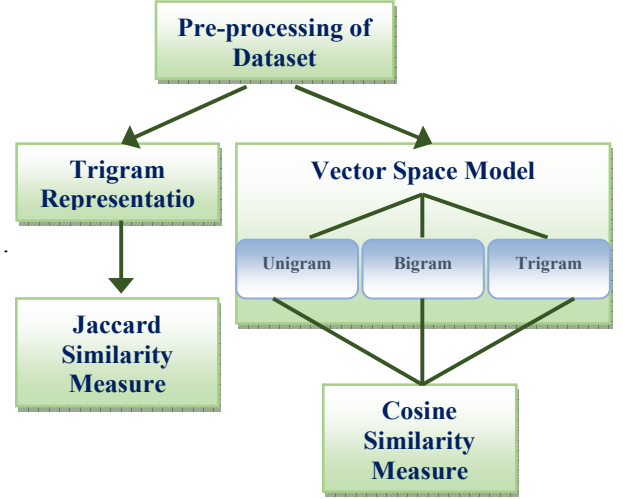

Fig. 1. System modules

A. *Vector Space Model with Cosine Similarity Measure*

Here, we have used vector space model using unigram, bigram and trigram approaches. Pre-processed dataset can be represented as a unigram vector, which consists of number of occurrences of a unigram in each assignment. Each assignment considered as a document. Document frequency for each term ($df_t$) was calculated by counting the number of documents which had the term t. Inverse document frequency for each term was computed using the following equation (1).

$$idf_t = 1 + log\, N / df_t \qquad (1)$$

$df_t$ - Document frequency for term t
$idf_t$ - Inverse document frequency for term t
$N$ - Total number of documents

The term, which appears in least number of documents contribute more to similarity than the terms which exists in many documents. Because, keywords in a document appear comparatively less than the common words and may occur in less number of documents. In contrast to this, the common terms may occur in many documents, which can show less impact on plagiarism detection. Therefore, the terms exist in less number of documents have been given more weight (inverse document frequency) than the terms which are found in many documents.

More number of occurrences of a term in a document is having more impact on plagiarism detection than less number of occurrences of a term. $tf_{t,d}$ -$idf_t$ weight vector is generated by product of term frequency and inverse document frequency as in equation (2).

$$W_{t,d} = tf_{t,d} \times idf_t \qquad (2)$$

$W_{t,d}$ - tf-idf weight vector
$tf_{t,d}$ - Frequency of a term in a document
$idf_t$ - Inverse document frequency for term t

Cosine similarity measure is computed for each pair of documents using the following equation (3).

$$Cos\,(d_i,\, d_j) = \Sigma\,(d_i,\, d_j) / (\sqrt{\Sigma d_i^2} \times \sqrt{\Sigma d_j^2}) \qquad (3)$$

$d_i, d_j$ -assignment pair.

### B. Trigram with Jaccard Similarity Measure

Here, the similarity score is computed using trigram sequence matching of Jaccard similarity measure for each pair of assignment using the given equation (4).

$$J(d_i, d_j) = |d_i \cap d_j| / |d_i \cup d_j| \quad (4)$$

$J$-Jaccard similarity
$d_i, d_j$ -assignment pair.

### C. Experimental Setup

The dataset was a collection of text based electronic assignments collected from the students and manually evaluated for plagiarism. Pre-processing of the dataset plays a significant role in the plagiarism detection computation. It involves eliminating delimiters, stop words, diagrams and pictures. The pre-processed dataset consists of collection of documents and represented as a bag of words of unigram, bigram and trigram. Vector space model with cosine similarity measure of unigram is applied to compute the percentage of plagiarism among assignments. Similarly, cosine similarity measure of bigram and trigram vectors are computed as well. Again the same dataset is represented as a list of sequence of trigram for each assignment and its each pair is compared and similarity score is computed using standard Jaccard similarity measure.

## IV. RESULTS AND DISCUSSION

The collected assignments from 40 students are computed with the given approach and the results are represented as follows.

The pre-processed assignments have been tested using three methods of vector space models and the maximum percentage of plagiarism among the combination is considered for each assignment.

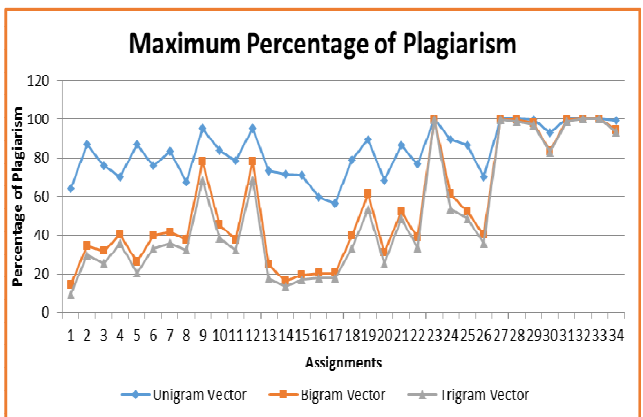

Fig. 2. Line chart for the comparison of plagiarism using unigram, bigram and trigram

Figure 2 shows the maximum percentage of plagiarism using cosine similarity measure with unigram, bigram and trigram vector. It elicits gradual decrease in percentage of plagiarism in each assignment regards to unigram, bigram and trigram vector. The plagiarism detection using unigram vector shows higher percentage than bigram and trigram vectors. Because, unigram vector space model does not represent consecutive sequence of terms in an assignment whereas bigram and trigram vector space consider two consecutive and three consecutive terms respectively. For example, "*Peter is quicker than Kerry*" and "*Kerry is quicker than Peter*" are two differently meaning sentences and they represent same vectors in unigram vector space model, where as they show different vectors for bigram and trigram. Even then, in the given graph, very highly plagiarized assignments such as 23, 27, 28, 29, 30, 31, 32, 33, 34 show approximately same percentage in all three methods due to the real and complete copying among them. Therefore, unigram vector space model is not an appropriate approach to detect copying of text based assignments and bigram or trigram approach can be more preferable.

Table I illustrates the control experiment we performed using three copies of an assignment labelled as A, B, and C and tested using three methods: 1) removing half part from copy A; 2) half part of the assignment is copied from A to B; and 3) the whole part of the assignment is copied from A to B. All pairs of the original assignments show 100% of plagiarism in all methods whereas it gradually decreases as 88%, 63% and 59% in unigram, bigram and trigram respectively while removing half part of A. In contrast to this, when half part and whole part of A is copied from A to B, there are no significant differences in the percentage of plagiarism due to the more number of common terms among them. Assignment C is maintained as unchanged in all three instances.

TABLE I. CONTROL EXPERIMENT USING THREE COPIES OF AN ASSIGNMENT

| Copies of Assignment Pairs | | Unigram | Bigram | Trigram |
|---|---|---|---|---|
| A | B | 100 | 100 | 100 |
| A | C | 100 | 100 | 100 |
| B | C | 100 | 100 | 100 |
| **Half Part of the Text is Removed from A** | | | | |
| A | B | 88 | 63 | 59 |
| A | C | 88 | 63 | 59 |
| B | C | 100 | 100 | 100 |
| **Half Part of the Text is Copied from A to B** | | | | |
| A | B | 99 | 96 | 94 |
| A | C | 100 | 100 | 100 |
| B | C | 99 | 96 | 94 |
| **Whole Part of the Text is Copied from A and Pasted to B** | | | | |
| A | B | 100 | 100 | 100 |
| A | C | 100 | 100 | 100 |
| B | C | 100 | 100 | 100 |

Table II represents a control experiment to find out the thresholds for each method. Here, five manually evaluated original assignments have been tested using the three methods to compute threshold values using the formula in Equation (5).

*Threshold=Maximum % of plagiarism + (Standard Deviation *4)*   (5)

The computed threshold values for plagiarism are 77%, 35% and 32% for unigram, bigram and trigram respectively.

TABLE II. CONTROL EXPERIMENT FOR THRESHOLD COMPUTATION

| Assignment Pairs | | Unigram | Bigram | Trigram |
|---|---|---|---|---|
| A | B | 64 | 7 | 1 |
| A | C | 60 | 7 | 1 |
| A | D | 60 | 6 | 1 |
| A | E | 60 | 4 | 1 |
| B | C | 62 | 10 | 7 |
| B | D | 63 | 7 | 2 |
| B | E | 59 | 4 | 1 |
| C | D | 58 | 5 | 1 |
| C | E | 65 | 18 | 14 |
| D | E | 54 | 4 | 1 |

Table III represents the comparison of manual detection and software detection according to the threshold. From the assignments, numbered 23 and 27-34 have been manually evaluated as completely copied and the others as partially copied. These labelled assignments are detected by our approach using our threshold values as well. The results match the manually evaluated labelled documents as shown in Table III.

TABLE III. COMPARISON OF PLAGIARISM DETECTION USING MANUAL AND SOFTWARE BASED APPROACH

| Assignment No | Manually Detected | Unigram | Bigram | Trigram |
|---|---|---|---|---|
| 1 | | | | |
| 2 | | √ | | |
| 3 | | | | |
| 4 | √ | | √ | √ |
| 5 | | √ | | |
| 6 | | | √ | √ |
| 7 | √ | √ | √ | √ |
| 8 | √ | | √ | √ |
| 9 | √ | √ | √ | √ |
| 10 | √ | √ | √ | √ |
| 11 | √ | √ | √ | √ |
| 12 | √ | √ | √ | √ |
| 13 | | | | |
| 14 | | | | |
| 15 | | | | |
| 16 | | | | |
| 17 | | | | |
| 18 | √ | √ | √ | √ |
| 19 | √ | √ | √ | √ |
| 20 | | | | |
| 21 | √ | √ | √ | √ |
| 22 | √ | | √ | √ |
| 23 | √√ | √√ | √√ | √√ |
| 24 | √ | √ | √ | √ |
| 25 | √ | √ | √ | √ |
| 26 | √ | | √ | √ |
| 27 | √√ | √√ | √√ | √√ |
| 28 | √√ | √√ | √√ | √√ |
| 29 | √√ | √√ | √√ | √√ |
| 30 | √√ | √√ | √√ | √√ |
| 31 | √√ | √√ | √√ | √√ |
| 32 | √√ | √√ | √√ | √√ |
| 33 | √√ | √√ | √√ | √√ |
| 34 | √√ | √√ | √√ | √√ |

√√-Complete copying
√-Partial copying

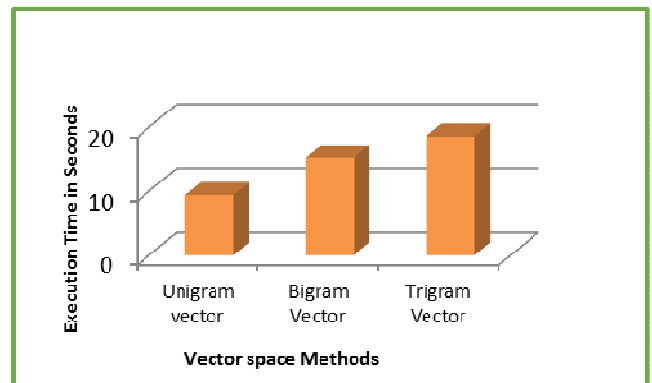

Fig. 3. Execution time of different vector space models

Figure 3 shows the execution time of unigram, bigram and trigram using vector space model as 9, 15, 18 seconds respectively. Trigram technique consumes comparatively more time than the others.

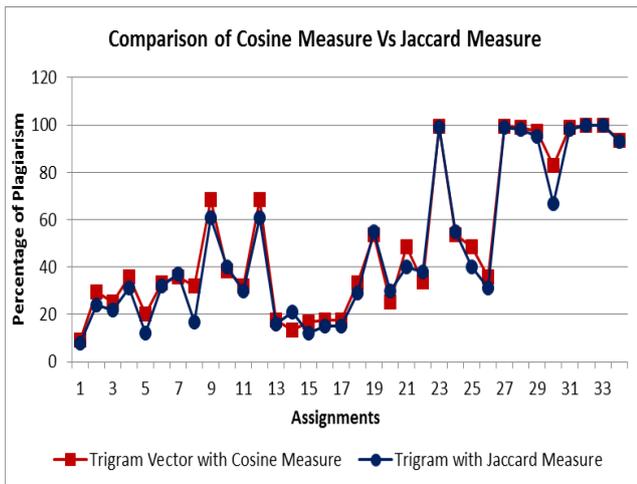

Fig. 4. Comparison of Cosine vs. Jaccard similarity measure

Figure 4 represents the maximum percentage of plagiarism of assignments using trigram with cosine similarity measure and trigram with Jaccard similarity measure. In the given chart cosine similarity in vector space model shows slightly higher percentage than the other plagiarism detection approach due to giving *tf-idf* weight for trigram in the vector space model.

## V. CONCLUSION

From the overall analysis, even though trigram consumes comparatively more time than the others, trigram method is more suitable for the detection of plagiarism in text documents using cosine similarity measure. Generally, the vector space model is used in information retrieval using query processing in massive pool of electronic resources. In this paper, we have proposed the vector space model using trigram as a suitable approach for plagiarism detection. In addition, cosine similarity measure shows slightly higher results than Jaccard similarity measure and therefore cosine similarity measure is more preferable than the other approach. This is because, the vector space model focuses on and provides more weights for terms that do not frequently exist in the dataset, whereas Jaccard similarity measure does not do this well.

The future work of this research is to focus on applying new approach to optimize percentage of plagiarism detection within minimum time when handling massive amount of assignments with higher document length.